\documentclass[preprint,12pt,3p]{elsarticle}

\usepackage{amssymb}
\usepackage{caption}
\usepackage{subcaption}
\usepackage{adjustbox}
\usepackage{url}
\usepackage{makecell}
\usepackage{float}
\usepackage{multirow}

\journal{Genomics, Proteomics \& Bioinformatics}

\begin{document}

\begin{frontmatter}

\title{CTCF Degradation Causes Increased Usage of Upstream Exons in Mouse Embryonic Stem Cells}


\author[label1,label2]{Boyi Yang, Ph.D.}   
\author[label1,label3]{Nabil Aounallah, M.D.}
\address[label1]{Harvard Medical School, Boston, MA}
\address[label2]{Brigham and Women's Hospital, Boston, MA}
\address[label3]{Massachusetts General Hospital, Boston, MA}




\begin{abstract}
Transcriptional repressor CTCF is an important regulator of chromatin 3D structure, facilitating the formation of topologically associating domains (TADs). However, its direct effects on gene regulation is less well understood. Here, we utilize previously published ChIP-seq and RNA-seq data to investigate the effects of CTCF on alternative splicing of genes with CTCF sites. We compared the amount of RNA-seq signals in exons upstream and downstream of binding sites following auxin-induced degradation of CTCF in mouse embryonic stem cells. We found that changes in gene expression following CTCF depletion were significant, with a general increase in the presence of upstream exons. We infer that a possible mechanism by which CTCF binding contributes to alternative splicing is by causing pauses in the transcription mechanism during which splicing elements are able to concurrently act on upstream exons already transcribed into RNA.
\end{abstract}

\end{frontmatter}


\section{INTRODUCTION}
\label{INTRODUCTION}

Recent Hi-C imaging and sequencing technology have elucidated the importance of 3-D chromatin structure and epigenetics in gene regulation \cite{nora2012spatial,rao20143d,lieberman2009comprehensive}. In addition to containing compartments of active, gene rich euchromatin and compartments of inactive, gene poor heterochromatin, chromatin is spatially partitioned into topologically associating domains (TADs) \cite{pombo2015three}. TADs are insulated regions of chromatin, where sequences within each respective region have more frequent interactions than with sequences in other regions of the genome. Borders of TADs are often marked by the presence of CCCTC-binding factor (CTCF) \cite{ji20163d}. CTCF is a highly conserved zinc finger protein that recognizes 50 base pair variant sequences throughout the genome \cite{ohlsson2001ctcf}. CTCF is thought to facilitate TAD formation by binding to two distant locations of DNA and then binding to itself, creating a loop of chromatin \cite{phillips2009ctcf,rao20143d}. Recent studies of CTCF have shown it is essential to loop formation, driving epigenetic forces of gene expression, but it is not essential to compartmentalization of chromatin into active and inactive regions \cite{nora2017targeted}. 

Although CTCF’s role in loop formation is well characterized, its role in gene regulation is less well understood. Previous studies have noted CTCF’s importance in gene regulation during development, showing that disruption of CTCF affects gene transcription in mouse oocytes \cite{wan2008maternal}. Other studies have shown that disruption of CTCF affects essential genetic pathways of cell proliferation, differentiation, and apoptosis \cite{torrano2005ctcf}. Recent studies have linked CTCF to alternative splicing of nearby genes. In mammalian CD45 genes, CTCF is thought to promote inclusion of exon 5 by pausing RNA polymerase II \cite{shukla2011ctcf}. Genome-wide, CTCF is thought to facilitate exon inclusion, or alternate exon usage, during RNA splicing by bringing exons in closer proximity to their promoters \cite{ruiz2017ctcf}. However, these studies remain limited in the scope of genes investigated or largely correlative, demanding a functional investigation of the effect of CTCF on alternative splicing.

Here, we used previously published ChIP-seq and mRNA-seq data from a CTCF knockdown mouse embryonic stem cell (mESC) model to examine the extent of CTCF dependent alternative splicing events  \cite{nora2017targeted}. Specifically, we compared exon usage in genes that contain a CTCF binding site in mESC lines tagged with an auxin-inducible degron (AID) for CTCF and untagged wild type, before and after treatment with auxin. We provide evidence that the presence of intragenic CTCF alters exon usage in a transcription direction dependent manner. We show that degradation of CTCF in an AID system results in a higher proportion of upstream exon usage in alternative splicing. These results support the direct role of CTCF in regulating alternative splicing during embryogenesis, and nominate a heritable epigenetic system that can be probed to better understand the pathology of alternative splicing driven diseases that arise during development.

\section{METHOD}
\label{METHOD}
\subsection{Data Retrieval}
\label{Data}
All data analyzed in this study were from the previously published Nora et al., 2017 paper \cite{nora2017targeted}. Expression levels for mRNA fragments were retrieved from the National Center for Biotechnology Gene Expression Omnibus (GSE98671). Experimental parameters and total reads were obtained from the supplements of Nora et al., 2017. CTCF ChIP-seq peak locations and magnitudes were provided by the Mirny Lab located at the Massachusetts Institute of Technology (mirnylab.mit.edu). Mouse genome mappings (NCBI37/mm9) were available from the University of California Santa Cruz (UCSC) Genome Browser.

\subsection{Identification and Ranking of Gene Bound CTCF Sites}
\label{Identification}
All calculations were done on R (Version 3.4.1) using tools from the Bioconductor project. First, the most prominent CTCF sites that were successfully degraded by auxin were isolated. From the 43,607 CTCF ChIP peaks in the untreated sample, 13,131 peaks remained or were not fully degraded in the treated sample. These 13,131 peaks were identified by genomic location using the findOverlaps function and subsequently removed from analysis. The 30,554 remaining peaks were then cross referenced with the known gene locations of the mm9 assembly to find CTCF sites that were located within protein coding sequences. This final pool of 16,665 peaks were ranked by peak magnitude and the highest five thousand were examined in this study. 

\subsection{Quantifying mRNA-Seq Reads}
\label{Quantifying}
For each of the 5,000 CTCF sites selected, the gene containing the CTCF site was found and the isoform with the most comprehensive selection of exons was selected. The locations of all exons upstream and downstream of the site were then identified. For each RNA-seq tag Density file, the signals in the exons upstream were summed. The resulting sum was divided by the total signal for the entire RNA-seq file and multiplied by the total number of reads in the experiment to estimate the number of reads in the upstream exons:
$$R_{mRNA} = \frac{\sum_{exon} Signal }{\sum_{total} Signal } R_{total} ,$$
where $R_{mRNA}$ is the mRNA-seq Reads and $R_{total}$ the total number of reads in our experiment. The same calculation was done for the downstream exons. Estimated reads were rounded to the nearest whole number and pooled with data from experimental replicates under the same conditions. 

\subsection{Statistical Analysis}
\label{Statistical}
The statistic used to compare distribution of isoforms around CTCF sites is the proportion ($P$) of reads upstream ($U$) compared to reads downstream ($D$), $P = \frac{U}{D}$. This statistic will hereafter be referred to as the proportion of a CTCF site. The final set of sites with valid proportion values consisted of 2,636 sites.

\section{RESULTS}
\label{RESULTS}
\subsection{Orientation and Shift in Proportions}
\label{Orientation}
\begin{figure}
	\centering
	\begin{subfigure}[b]{0.3\textwidth}
		\centering
		\includegraphics[width=\textwidth]{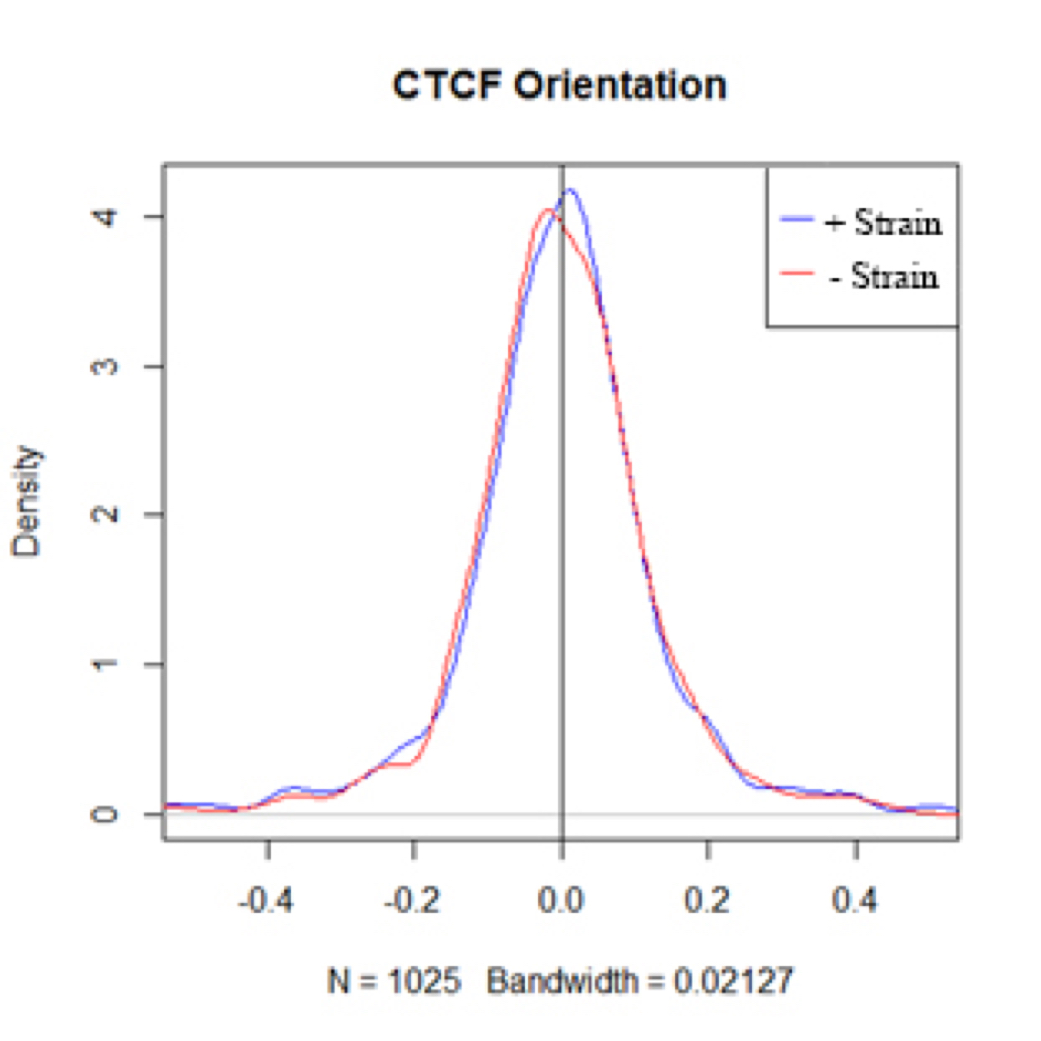}
		\caption{}
		\label{fig1a}
	\end{subfigure}
	\begin{subfigure}[b]{0.3\textwidth}
		\centering
		\includegraphics[width=\textwidth]{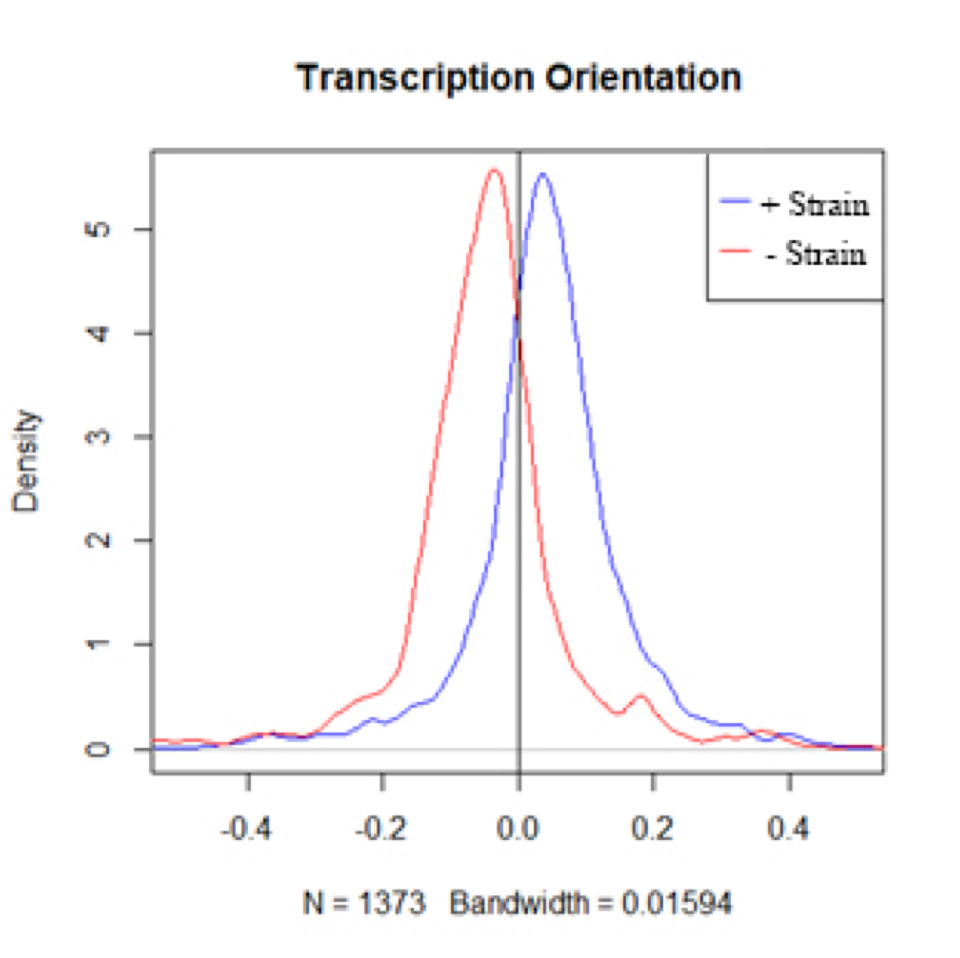}
		\caption{}
		\label{fig1b}
	\end{subfigure}
	\begin{subfigure}[b]{0.3\textwidth}
		\centering
		\includegraphics[width=\textwidth]{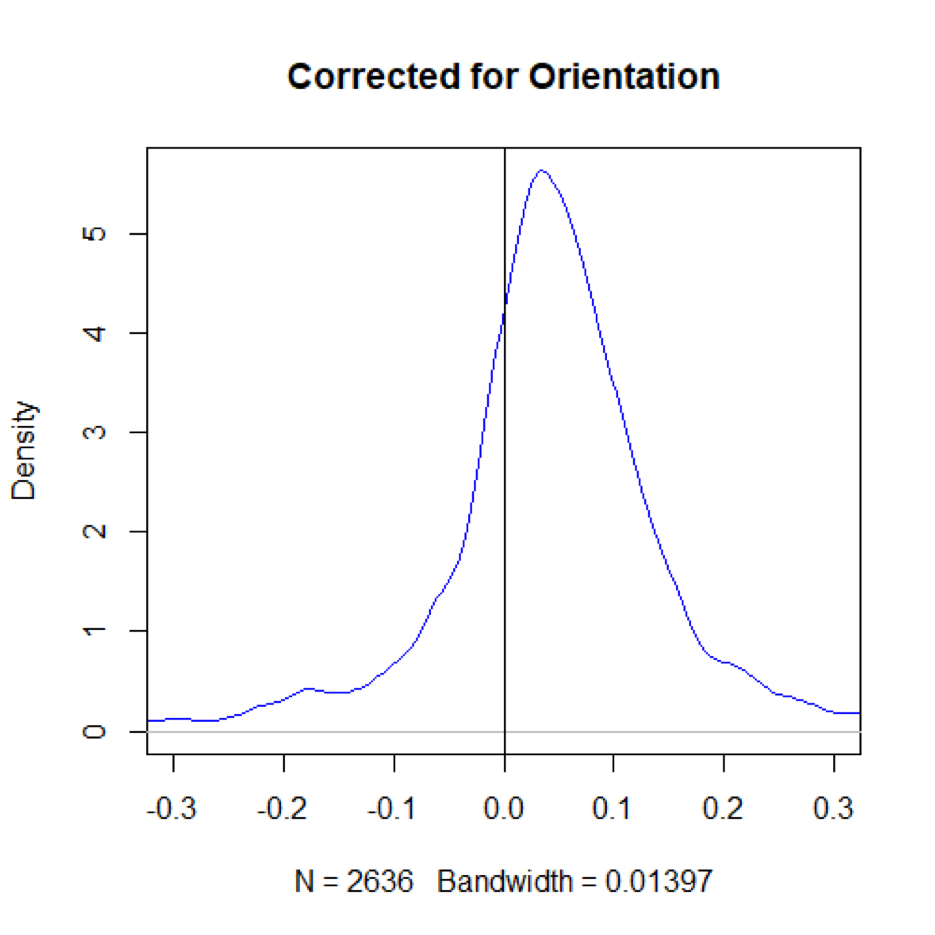}
		\caption{}
		\label{fig1c}
	\end{subfigure}
	\caption{Log change in proportions in CTCF-AID tagged cells from untreated to auxin 2 days ($N=2636$). (a) Distributions grouped by CTCF orientation overlap and show no significant difference. (b) Distributions grouped by transcription orientation are mirrored and show significant difference ($p.value=7E – 63$). (c) When proportions are recalculated to account for transcriptional orientation, log change in proportions show significant deviation from zero ($p.value=2E – 63$).}
	\label{fig1}
\end{figure}

Before the influence of CTCF on alternative splicing can be examined, directionality effects due to CTCF and transcriptional direction have to be accounted for. Effects of various experimental conditions were quantified by dividing the proportion after treatment by the proportion before treatment. Kernel Density plots for the log change in proportions are plotted in Figure \ref{fig1}. As Figure \ref{fig1a} shows, changes in log expression ratio around the CTCF site does not depend on CTCF orientation. On the other hand, Figure \ref{fig1b} shows that changes in log expression ratios are symmetric with respect to transcriptional direction. A two-sample t-test shows significant difference ($p.value = 2E-63$) between the distributions. Thus, comparisons must be made with respect to transcription orientation, with upstream of a CTCF site being defined as transcriptionally upstream and downstream as transcriptionally downstream. Once corrected for transcriptional direction, change in log expression ratio is positive (Figure \ref{fig1c}).

\subsection{Contingency Tables and Tests for Significance}
\label{Contingency}
To get an understanding of specific changes in isoform distribution and evaluate the significance of the change at specific CTCF sites, contingency tables were built for each site. Observations consist of the number of fragments detected upstream and downstream of the CTCF site with Fisher’s exact test conducted to evaluate the difference. Fisher’s exact test was preferred over chi-squared test because reads of mRNA fragment tend to be skewed and vary wildly between very large and very small counts. Multiple testing correction was performed using Bonferroni correction, resulting in a more conservative alpha for significance testing, $\alpha=0.05/2636=1.9E-5$.
Given the noisiness of the data, the reduced power owing to increased conservativeness of the test and FDR control were acceptable.
Three contingency tables were constructed for each site to evaluate the influence of CTCF degradation on alternative splicing. The parameters for the tests are expressed in Table \ref{tab1}. 

\begin{table}[H]
	\centering
	\begin{adjustbox}{width=1\textwidth}
		\begin{tabular}{|l|l|l|l|}
			\hline
			\textbf{Test}   & \textbf{Sample}  & \textbf{Comparison} & \textbf{Observed}\\ 
			\hline
			1  &  CTCF-AID tagged cells  & Untreated vs auxin 2 days  & \thead{Distribution of fragments \\upstream and downstream\\ of the CTCF site}\\ 
			\hline
			2  &  Wildtype untagged cells & Untreated vs auxin 2 days& \thead{Distribution of fragments \\upstream and downstream\\ of the CTCF site}\\
			\hline
			3  &  Untreated tagged and untagged cells  & CTCF-AID tagged vs wildtype untagged& \thead{Distribution of fragments \\upstream and downstream\\ of the CTCF site}\\
			\hline 			
		\end{tabular}
	\end{adjustbox}
	\caption{Description of tests conducted on contingency tables}
	\label{tab1}
\end{table}

\subsection{Changes in Proportions in CTCF Bound Genes}
\begin{figure}
	\centering
	\begin{subfigure}[b]{0.3\textwidth}
		\centering
		\includegraphics[width=\textwidth]{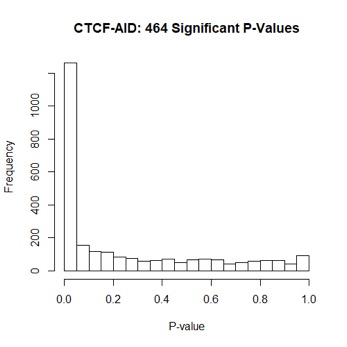}
		\caption{}
		\label{fig2a}
	\end{subfigure}
	\begin{subfigure}[b]{0.3\textwidth}
		\centering
		\includegraphics[width=\textwidth]{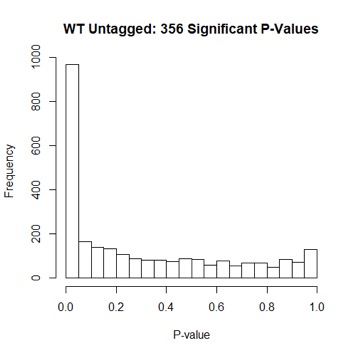}
		\caption{}
		\label{fig2b}
	\end{subfigure}
	\begin{subfigure}[b]{0.3\textwidth}
		\centering
		\includegraphics[width=\textwidth]{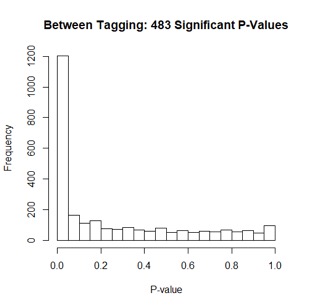}
		\caption{}
		\label{fig2c}
	\end{subfigure}
	\begin{subfigure}[b]{0.3\textwidth}
	\centering
	\includegraphics[width=\textwidth]{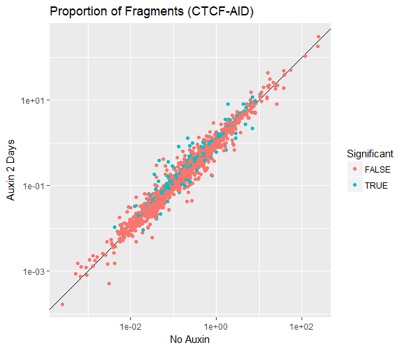}
	\caption{}
	\label{fig2d}
	\end{subfigure}
	\begin{subfigure}[b]{0.3\textwidth}
	\centering
	\includegraphics[width=\textwidth]{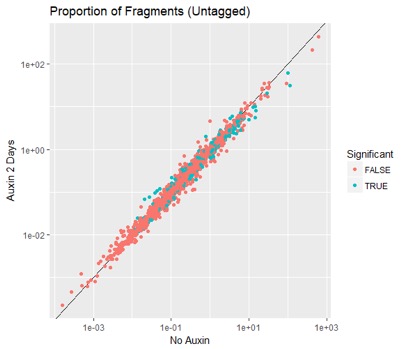}
	\caption{}
	\label{fig2e}
	\end{subfigure}
	\begin{subfigure}[b]{0.3\textwidth}
	\centering
	\includegraphics[width=\textwidth]{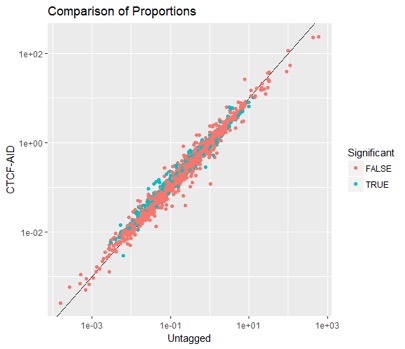}
	\caption{}
	\label{fig2f}
\end{subfigure}
	\caption{Factors that Affect Alternative Splicing. (a – c) Distribution of p-values from Fisher’s exact test of factors influencing splicing. (a) Test 1 evaluates the effect of auxin on CTCF-AID tagged cells. (b) Test 2 evaluates the effect of auxin on wildtype untagged cells. (c) Test 3 evaluates the effect of CTCF-AID tagging. (d – f) scatter plots mapping proportions in control and experiment and color coded by significance of p-values. (d, e) Untreated vs auxin 2 days on tagged and untagged WT cells. (f) Untagged WT vs CTCF-AID tagged untreated cells. }
	\label{fig2}
\end{figure}

The distribution of p-values show anti-conservative trends for all three tests (Figure \ref{fig2}), suggesting that the alternative hypothesis of equal exon usage in genes with CTCF binding sites may be true for some genes. In CTCF-AID tagged cells, treatment with auxin resulted in significant changes in proportions at 464 CTCF sites (Figure \ref{fig2a}). Surprisingly, 356 sites in wildtype untagged cells also showed significant change after auxin treatment (Figure \ref{fig1b}) even though treatment wouldn’t result in CTCF depletion. Moreover, comparing untreated CTCF-AID to untreated untagged cells shows 483 significant sites (Figure \ref{fig2c}). 

\begin{figure}
	\centering
	\begin{subfigure}[b]{0.45\textwidth}
		\centering
		\includegraphics[width=\textwidth]{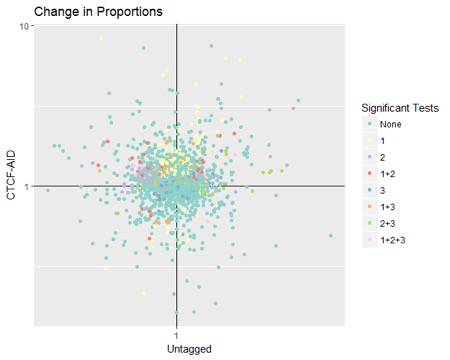}
		\caption{}
		\label{fig3a}
	\end{subfigure}
	\begin{subfigure}[b]{0.4\textwidth}
		\centering
		\includegraphics[width=\textwidth]{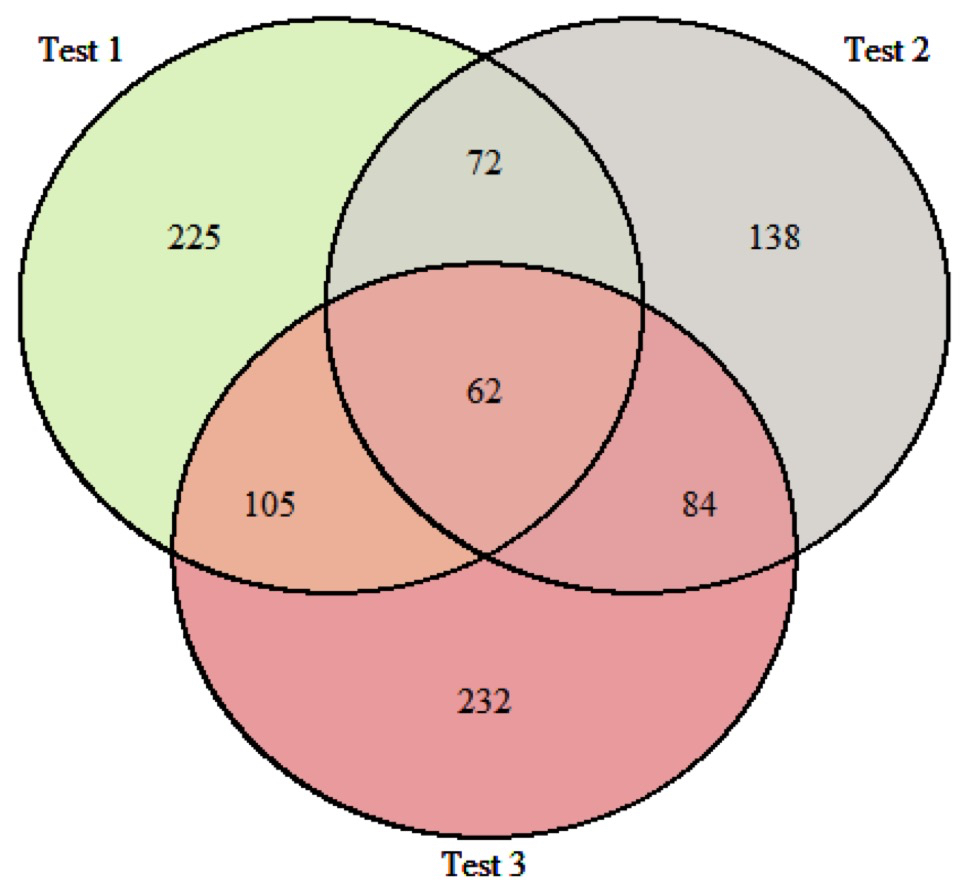}
		\caption{}
		\label{fig3b}
	\end{subfigure}
	\caption{CTCF sites grouped by tests they are significant for. (a) Scatter plot mapping log change in proportion in wildtype untagged cells to log change in CTCF-AID tagged cells. (b) Venn diagram showing overlaps in test significance. There were 1718 sites which were not significant for any of the tests.}
	\label{fig3}
\end{figure}

A large magnitude of change in proportion isn’t always significant while a large number of site that showed moderate proportion changes were significant. Significant sites showed both positive and negative change in proportions (Figure \ref{fig3a}). Nonsignificant points showing large changes had relatively small numbers of reads, making it possible for small variations to show a large magnitude change but still give high p-values; meanwhile, sites showing significance close to the center have a large number of reads (Supplementary Table Sites). A number of sites overlap in the tests for which they are significant (Figure \ref{fig3b}). There are more overlaps than expected for such stringent selection, suggesting that there may be common mechanisms causing these sites to display greater variation. 

\subsection{Distribution of Proportion Change}
\label{Distribution}
\begin{figure}
	\centering
	\begin{subfigure}[b]{0.8\textwidth}
		\centering
		\includegraphics[width=\textwidth]{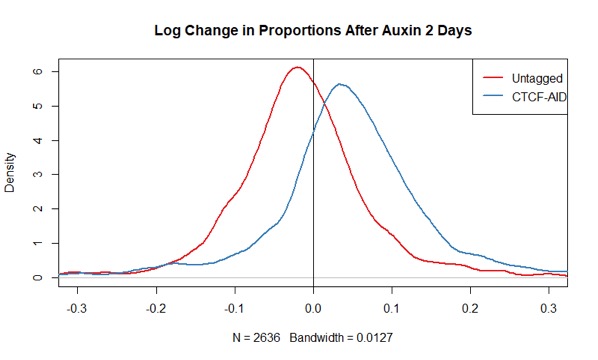}
		\caption{}
		\label{fig4a}
	\end{subfigure}

	\begin{subfigure}[b]{0.8\textwidth}
		\centering
		\includegraphics[width=\textwidth]{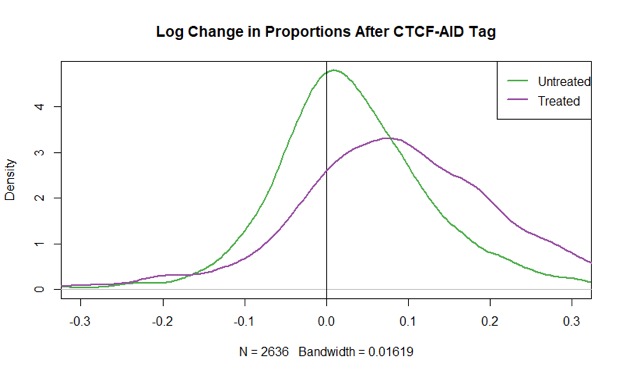}
		\caption{}
		\label{fig4b}
	\end{subfigure}
	\caption{Log Change in proportions across treatments. (a) Comparing untreated to auxin 2 days. Untagged cells show decrease with $p.value=3E-16$. Tagged cells show increase with $p.value=2E-63$. (b) Comparing wildtype untagged to CTCF-AID tagged. Untreated cells show increase with $p.value=2E-51$. Treated cells show increase with $p.value=5E-171$.}
	\label{fig4}
\end{figure}

Assuming that all CTCF sites are in one population and impacted similarly, the distribution of proportion changes can be viewed as a whole to analyze how treatment affected splicing. Exposure to auxin for 2 days caused proportions to significantly increase in CTCF-AID tagged cells and to significantly decrease in wild-type untagged cells (Figure \ref{fig4a}). As expected, treated tagged cells showed significant increase in proportions compared to treated untagged cells. Surprisingly, untreated tagged cells also showed significant increases over untreated untagged cells. However, it is of note that these changes were smaller in magnitude (Figure \ref{fig4b}). 

\section{DISCUSSION}
\label{DISCUSSION}
While CTCF has been noted as a key player in shaping the 3D structure of chromatin, its direct effects on gene regulation, and in particular alternative splicing, is less well characterized. As noted, this study investigated the direct effects of CTCF on alternative exon usage in mouse embryonic stem cells. Using previously published ChIP-seq and RNA-seq data from Nora et al., 2017, we investigated alternative splicing in genes containing a CTCF binding site. We quantified and compared the changes in alternative exon usage after the CTCF was removed from a CTCF bound gene was degraded. 

We found that degrading CTCF using auxin in CTCF-AID tagged cells resulted in an increase in the proportion of upstream fragments used in final mRNA transcripts. This finding supports Shukla et al, 2011’s hypothesis that CTCF binding to DNA blocks transcription proteins and causes pauses in mRNA transcription. Splicing occurs concurrently with transcription, and thus when transcription is paused due to CTCF, splicing elements are able to act upon RNA upstream that were already transcribed with greater frequency. Depletion of CTCF likely prevents these pauses in transcription, giving splicing elements less opportunity to act on exons upstream of CTCF sites. Here, we provide evidence for this mechanism by showing that CTCF depletion results in greater upstream exon usage in mRNA formation. 

Although these findings seem promising, it should be noted that significant differences in the proportions of exon usage were also observed in control cases that should exhibit none. In particular, there were 384 genes that exhibited a significant change in alternative splicing in the wild type, untagged cells after treatment with auxin. When comparing wild type, untagged cells with CTCF-AID tagged cells without the presence of auxin, we found 483 CTCF bound genes that exhibited significantly different alternative splicing among these two conditions. Although the magnitudes of the changes were smaller than those between the experimental condition and control, these changes are alarming as they suggest that the AID tagging method itself may cause changes in gene expression. One explanation is that the tagging process affected the expression levels of genes related to splicing factors and transcription controls. Another is that although ChIP-seq shows tagged CTCF still bound to DNA, the binding efficiency may be impacted to an extent that differences in splicing may be observed. What may be even harder to explain is why untagged cells showed lower proportions after exposure to auxin. Untagged cells did not have their CTCF degraded, showed no difference in gene expression levels, and did not suffer from the same cytotoxic effects that tagged cells displayed. Nonetheless, untagged cells showed an anti-conservative distribution of p-values and a significantly negative change in proportion. These observations should be examined further, or they may undermine the conclusions made above.

The effect of CTCF binding on alternative splicing in mouse embryonic stem cells is apparent in this study. We showed a number of genes that exhibit a change in upstream exon usage after depletion of CTCF, suggesting a functional role of CTCF in determining alternatively spliced mRNA transcripts. As noted by Li et al., 2016, alternative splicing and the resulting isoforms have great impact not only on biodiversity and genetic variation but also on disease \cite{li2016rna}. Understanding the mechanisms behind alternative splicing can provide insight into the pathology of diseases such as developmental disorders and cancers. This has already been hinted by Filipova et al., 1998 when the authors associated CTCF binding with deletions resulting in breast and prostate cancers \cite{filippova1998widely}. Perhaps more exciting, understanding the decision-making machinery behind alternative splicing can expose potential vulnerabilities in alternative splicing driven mechanism and inform potential targets for therapy. Future work should focus on characterizing the types of genes and pathways affected by CTCF mediating alternative splicing. 

\bibliographystyle{elsarticle-num}

\bibliography{CTCF}

\end{document}